\documentclass[conference, anonymous]{IEEEtran}
\IEEEoverridecommandlockouts
\usepackage{cite}
\usepackage{amsmath,amssymb,amsfonts}
\usepackage{algorithmic}
\usepackage{graphicx}
\usepackage{textcomp}
\usepackage{xcolor}
\usepackage{cite}
\usepackage{float}
\usepackage[hyphens]{url}
\def\BibTeX{{\rm B\kern-.05em{\sc i\kern-.025em b}\kern-.08em
    T\kern-.1667em\lower.7ex\hbox{E}\kern-.125emX}}

\begin{document}

\title{Experimental Shake Gesture Detection API for Apple Watch}

\newcommand{\pucsp}{Pontifícia Universidade Católica de São Paulo (PUC-SP)}
\newcommand{\dolemes}{David de Oliveira Lemes}
\newcommand{\ezequiel}{Ezequiel França dos Santos}

\author{\IEEEauthorblockN{\ezequiel}
\IEEEauthorblockA{\textit{Phd Student in Digital Games Development} \\
\textit{IADE — Universidade Europeia}\\
Lisbon, Portugal \\
ezequield@acm.org}
}

\maketitle

\begin{abstract}
In this paper, we present the WatchShaker project. The project involves an experimental API that detects the Apple Watch's shake gesture—a surprisingly absent natively feature. Through a simple heuristic leveraging the Apple Watch's accelerometer data, the API discerns not just the occurrence of shake gestures but also their direction, enhancing the interactivity potential of the device. Despite the project's simplicity and lack of formal testing, it has garnered significant attention, indicating a genuine interest and need within the developer community for such functionality. The WatchShaker project exemplifies how a minimalistic approach can yield a practical and impactful tool in wearable technology, providing a springboard for further research and development in intuitive gesture recognition.
\end{abstract}


\section{Introduction}

Wearable technology has grown widely, with developers leveraging motion sensors for various gesture recognition applications. These applications span across various sectors, including entertainment systems \cite{7030602} \cite{10.1109/CCNC.2016.7444820}, health systems \cite{10.1145/2594368.2594379}, and more, showcasing the technology's versatility and broad impact \cite{6839278}.

Despite its sophisticated ecosystem, the Apple Watch presents unexplored functionalities and user interaction areas. Notably, it lacks native support for shake gesture detection, a feature prevalent in its smartphone counterparts \cite{appleUIEventEventSubtypemotionShakeApple}\cite{Freeman2015-nt}. The "WatchShaker" project addresses this gap by developing an API for shake gesture detection, aiming to enrich the interaction capabilities of the Apple Watch. This paper explores the development process of WatchShaker, evaluates its implications for Apple Watch users, and discusses its contribution to the broader field of Human-Computer Interaction (HCI) and wearable technology. Integrating this intuitive gesture, the WatchShaker enhances the usability of the Apple Watch and opens new avenues for applications and user experiences.

\subsection{Background and Motivation}
The evolution of smartwatches has significantly transformed them from simple time-keeping devices to integral components of our digital lives \cite{10.3390/mti2030038}. As one of the latest developments in the evolution of information technology, smartwatches offer users a remarkable level of convenience, swiftly and discreetly delivering timely information with minimal interference or intrusion compared with smartphones and other mobile devices \cite{10.3390/mti2030038}. However, the interaction modalities with these devices are still evolving, and the Apple Watch, a frontrunner in this space, lacks specific gesture control standards in other smart devices \cite{10.1145/2702613.2725441}. The WatchShaker project was created as an experiment to enhance the functionality of the Apple Watch and make it more user-friendly. This project aims to make the device more intuitive and in sync with other smart devices.


\subsection{Objectives of WatchShaker}
The primary objective of the WatchShaker project is to introduce a reliable and responsive shake gesture detection API for the Apple Watch. This API aims to:
\begin{itemize}
    \item Provide an easy-to-integrate solution for app developers to incorporate shake gestures into their applications.
    \item Enhance the user experience by enabling a natural and quick way to interact with the watch.
    \item Explore new application design and functionality possibilities that leverage shake gestures.
    \item Contribute to the research and development in the field of HCI, particularly in wearable technology.
\end{itemize}

\section{Literature Review}
\subsection{Gesture Recognition in Wearables}
Recent advancements in wearable technology have highlighted the innovative potential of intuitive gesture recognition. The study "Enabling Hand Gesture Customization on Wrist-Worn Devices" \cite{10.1145/3491102.3501904} represents a significant contribution in this field, demonstrating the feasibility and effectiveness of customizable gesture recognition systems in wearables. Additionally, research on "Gesture Recognition Using Inertial Sensors with 1D Convolutional Neural Network" \cite{10140222} further validates the application of advanced sensor technologies and computational techniques in enhancing gesture recognition in wearable devices. The framework proposed in these studies achieves high accuracy and user adaptability, marking a notable advancement in the wearable technology landscape.

\subsection{API Development for Apple Watch}
While the development of APIs for the Apple Watch, as discussed in various works including \cite{10.1145/3491102.3501904}, has primarily focused on health and fitness applications, these studies reveal a broader potential for gesture control in enhancing user experience. They demonstrate advanced gesture recognition systems' technical feasibility and user-centric benefits, laying the groundwork for further explorations in this domain.

\subsection{Gap in Existing Research}
Despite these advancements, there is still an opportunity in gesture recognition for the Apple Watch, particularly concerning shake gesture detection. Papers like "Enabling hand gesture customization on wrist-worn devices" \cite{10.1145/3491102.3501904} and "Interaction with Smartwatches Using Gesture Recognition: A Systematic Literature Review" \cite{8029952} provide comprehensive insights into gesture recognition capabilities but do not explicitly address shake gestures. This omission highlights a specific gap in the current research landscape, underscoring the unique opportunity the WatchShaker project aims to explore and fulfill.

\section{Methodology}

The approach to developing the WatchShaker API was a hybrid methodology that combined traditional software development with Design Science Research (DSR) \cite{vom2020introduction}. This approach facilitated the creation of a functional API and framed it within a research context emphasizing practical problem-solving and artifact evaluation.

\subsection{Problem Identification}
Initially, the need for enhanced gesture recognition on the Apple Watch was identified, explicitly focusing on the absence of a native shake gesture detection feature. This need formed the basis for the WatchShaker project, aligning with the DSR's emphasis on addressing real-world problems.

\subsection{Design and Development}
The design phase concentrated on the API's architecture, creating key components such as \texttt{Shake}, \texttt{ShakeCoordinates}, \texttt{ShakeDirection}, and \texttt{ShakeSensibility}. These elements' design helps capture and interpret the shake gesture accurately and responsively, reflecting the iterative and user-centric nature of DSR.

\subsection{Implementation}
The developed core logic for shake detection uses accelerometer data from the Apple Watch. The following heuristic forms the basis of the gesture recognition algorithm, embodying the practical artifact creation aspect of DSR:

Let \( \textit{ShakeDetected} \) be a boolean variable defined by the following conditional expression:

\[
\text{\textit{ShakeDetected}} = 
\begin{cases}
\text{True} & \text{if } \max(|a_x|, |a_y|) > \theta \text{ and } t_{\Delta} > \tau, \\
\text{False} & \text{otherwise}.
\end{cases}
\]

Where:
\begin{itemize}
    \item \( a_x \) and \( a_y \) represent the acceleration values along the x-axis and y-axis, respectively.
    \item \( \theta \) denotes the predetermined threshold value for the shake sensitivity.
    \item \( t_{\Delta} \) signifies the elapsed time interval between the current observation and the most recent shake event.
    \item \( \tau \) is the designated delay period required for a subsequent shake detection to be considered valid.
\end{itemize}

This formulation allows for the discernment of significant movement events based on acceleration parameters and temporal constraints, thereby enhancing the accuracy and efficacy of motion detection algorithms. This heuristic formed the basis for the gesture recognition algorithm, focusing on ensuring accurate and responsive detection of shake gestures on the Apple Watch.

\subsection{Directional Shake Recognition}

\begin{figure}[H]
 \centering
 \includegraphics[width=0.35\textwidth]{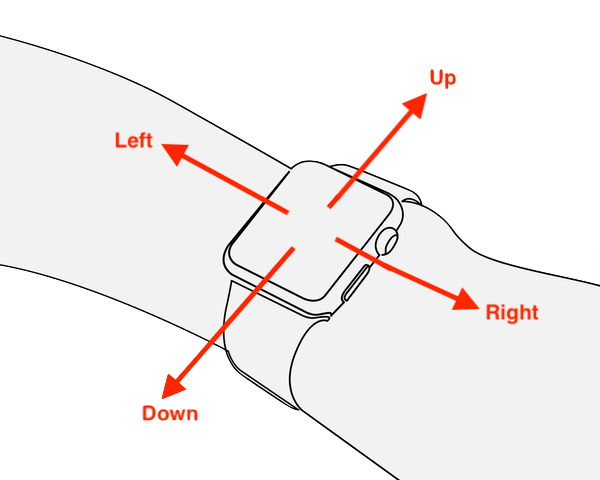}
\caption{Diagram illustrating the WatchShaker API's directional criteria for determining shake directions.}
 \label{fig:direction}
\end{figure}

The WatchShaker API extends its functionality beyond mere shake detection by incorporating Directional Shake Recognition (DSR). This feature enables the API to discern the direction of a shake gesture—be it up, down, left, or right. The directionality of the shake is determined through a heuristic analysis of the accelerometer data, as illustrated in Figure \ref{fig:direction}, thereby augmenting the API's utility and demonstrating its commitment to functional innovation.

Let \( \textit{ShakeDirection} \) be a variable that returns the shake direction, defined by the following conditional expression:

\[
\text{\textit{ShakeDirection}} = 
\begin{cases} 
\text{Right} & \text{if } |a_x| > |a_y| \text{ and } a_x > 0, \\
\text{Left} & \text{if } |a_x| > |a_y| \text{ and } a_x < 0, \\
\text{Down} & \text{if } |a_x| < |a_y| \text{ and } a_y > 0, \\
\text{Up} & \text{if } |a_x| < |a_y| \text{ and } a_y < 0, \\
\text{Unknown} & \text{otherwise}.
\end{cases}
\]

Where \( a_x \) and \( a_y \) denote the acceleration values in the x and y directions, respectively. This approach endows the API with the ability to provide a directionally informed response to shake gestures, offering developers enhanced control for application interactions. 

The API method \texttt{didShakeWithDirection} is designed to deliver this directional information, facilitating the implementation of directionally responsive gesture controls within various applications.

\subsection{Artifact Evaluation}
Due to resource constraints, formal testing was not feasible; however, the API underwent extensive personal and community evaluations. Feedback from platforms like GitHub and Stack Overflow provided an informal yet valuable means of assessing the artifact's effectiveness and usability, aligning with the DSR's evaluation criteria.

\subsection{Reflection and Learning}
The project's development process involved continuous reflection and adaptation, essential aspects of DSR. This phase included learning from the successes and limitations encountered and informing future iterations of the API.

\subsection{Contribution to Knowledge}
The WatchShaker project contributes to the HCI and wearable technology body of knowledge by demonstrating a practical and innovative solution for gesture recognition. The blend of traditional development with DSR highlights the project's role in advancing applied technology research.

This hybrid methodology underscores the WatchShaker API's practical development and assessment within a research framework, emphasizing the project's contribution to technology and applied research.

\subsection{Testing and Refinement}
The API was then subjected to a series of tests to evaluate its performance and accuracy. It involved automated testing and real-world usage scenarios to ensure the API could reliably detect shake gestures under various conditions.

\subsection{Finalization}
The final phase included optimizing the API for performance, preparing documentation for ease of use and integration by other developers, and making final adjustments based on testing feedback.

\subsection{Implementation}
The implementation phase involved the actual coding of the API in Swift. Key functionalities included:
\begin{itemize}
    \item The \texttt{WatchShaker} class serves as the primary interface for the shake service, managing motion sensor data and determining when a shake event occurred.
    \item Utilization of the CoreMotion framework to access accelerometer data, crucial for detecting shake gestures.
    \item Algorithms for interpreting accelerometer data to detect shake gestures, considering factors like sensibility and shake direction.
\end{itemize}
Iterative development and testing addressed challenges in interpreting sensor data and reducing false positives.

\subsection{Testing and Refinement}
Shake detection accuracy and responsiveness were validated by simulating shakes and assessing the API's ability to identify them under various conditions.

\subsection{Finalization}
The final stage involved optimizing the code for efficiency, preparing documentation to assist developers in integrating the API, and ensuring compatibility with different Apple Watch operating system versions.

This methodology led to the development of a versatile and user-friendly shake gesture detection API tailored for the unique environment of the Apple Watch.

\section{Testing and Results}
The WatchShaker API demonstrated its effectiveness through various recognition and personal testing forms, though not through formal academic or industry-standard methods.

\subsection{Recognition in Developer Community}
\begin{itemize}
    \item \textbf{GitHub Engagement}: The project's reception on GitHub, with over 217 stars and contributions from more than seven developers, reflects its relevance and usefulness.
    \item \textbf{Online Discussions}: Discussions and questions on platforms like \textit{Stack Overflow}\footnote{https://stackoverflow.com/questions/62395087/watchos-gyro-calibration-interrupted} provide real-world insights into the API's application and user challenges.
    \item \textbf{Feature in Developer Newsletter}: The API was featured in a developer newsletter \textit{Swift News Issue 118}
    \footnote{https://swiftnews.curated.co/issues/118}, highlighting its innovation and utility in the wider developer community.
\end{itemize}

\subsection{Personal Testing}
\begin{itemize}
    \item \textbf{Functional Testing}: API functionality, accuracy, and responsiveness, through personal tests, which aided in iterative refinement for better performance, were evaluated.
    \item \textbf{User Experience Evaluation}: Through hands-on use, the API's ease of integration and user interaction, ensuring that it met the intended goals of enhancing the Apple Watch's usability, were evaluated.
\end{itemize}

\subsection{Implications of Results}
Community engagement, developer newsletter recognition, and personal testing results testify to the API's effectiveness. These results indicate the API's technical capabilities, acceptance, and potential impact on the developer community.

In conclusion, the WatchShaker API, while not tested through formal industry standards, has shown considerable promise and effectiveness through community engagement and practical application, underscoring its potential for broader adoption and further development.

\section{Discussion}
\subsection{Simplicity and Effectiveness of the Approach}
The WatchShaker project stands out for its straightforward and unpretentious approach to solving a specific problem - detecting shake gestures on the Apple Watch. Using a basic heuristic exemplifies the project's commitment to simplicity while effectively fulfilling its intended purpose.

\subsection{Community Response and Practical Impact}
Despite its simplicity, the project has resonated well within the developer community, as evidenced by the engagement on GitHub and discussions on platforms like Stack Overflow. The feature in a developer newsletter further highlights its practical utility. This reception suggests that even simple ideas, when executed well, can have a significant impact.

\subsection{Real-World Application and Limitations}
The real-world application of the WatchShaker API, while not extensively tested through formal methods, has shown promising results. User discussions point towards its functional usability, highlighting areas where further refinement could be beneficial. This feedback is invaluable for future iterations of the project.

\subsection{Comparison with Other Technologies}
Compared to more complex solutions, the WatchShaker API offers an accessible alternative for gesture recognition on the Apple Watch. Its straightforward approach may only encompass some of the nuances of advanced gesture detection systems, but it offers a viable solution for many practical applications.

\subsection{Reflection on the Project and Future Directions}
The WatchShaker project demonstrates that simple ideas can be both practical and well-received. It opens the door for further exploration into similar straightforward solutions in wearable technology, proving that sometimes, less can be more.

While our study has focused on a pioneering approach to shake gesture detection for the Apple Watch, we acknowledge certain limitations in our current methodology, particularly in formal testing. Due to the innovative and exploratory nature of the WatchShaker project, our initial focus was on demonstrating the feasibility and garnering community feedback rather than on extensive empirical testing.

In deduction, through its modest yet practical design, the project contributes to the broader field of wearable technology, reminding us of the value and impact that simplicity can bring to technological innovation.

\section{Conclusion}

The WatchShaker\cite{https://doi.org/10.5281/zenodo.8054488} project exemplifies the principle that simplicity can drive innovation in technology. By focusing on a specific problem—detecting shake gestures on the Apple Watch—the project aimed to provide a straightforward and effective solution without overhauling the entire concept of gesture recognition. 

Despite its unassuming approach, the project garnered recognition and utility within the developer community, as evidenced by its GitHub stars and the discussions it sparked amongst peers. The practical application of the WatchShaker API has demonstrated promising results, with personal testing showing the API's reliability. Although this testing did not adhere to formal industry standards, the developer community's positive feedback and the recognition of developer newsletters provided empirical evidence of the API's value.  

As wearable technology continues to evolve, the WatchShaker project is a valuable reference point for developers seeking to implement intuitive gesture controls. The project's journey from a mere concept to a functional tool that enhances user interaction with the Apple Watch offers insights into the potential for further innovation in wearable technology, driven by community collaboration and a commitment to user-centric design.  

Future work on the WatchShaker project can build on the foundation laid by this initial implementation, expanding its functionality and refining its capabilities. The project's success underscores the potential for meaningful technological advancements through a simple and focused approach, inspiring further innovation in wearable technology.

\bibliography{sources}{}
\bibliographystyle{plain}

\end{document}